# Spreading of EGF Receptor Activity into EGF-free Regions and Molecular Therapies of Cancer

Garrit Jentsch* [1] and Reiner Kree [2]

(1) Department of Molecular, Cellular and Developmental Biology, Yale University, 266 Whitney Ave, New Haven, CT 06520, USA

(2) Inst. f. Theoret. Physik, Georg-August Universität Göttingen, Friedrich-Hund-Platz 1, D-37077 Göttingen, Germany

(*) To whom correspondence should be addressed. email: garrit.jentsch@yale.edu

**Abstract** The primary activation of the epidermal growth factor receptor (EGFR) has become a prominent target for molecular therapies against several forms of cancer. But despite considerable progress during the last years, many of its aspects remain poorly understood. Experiments on lateral spreading of receptor activity into ligand-free regions challenge the current standard models of EGFR activation. Here, we propose and study a theoretical model, which explains spreading into ligand-free regions without introducing any new, unknown kinetic parameters. The model exhibits bistability of activity, induced by a generic reaction mechanism, which consists of activation via dimerization and deactivation via a Michaelis-Menten reaction. It possesses slow propagating front solutions and faster initial transients. We analyze relevant experiments and find that they are in quantitative accordance with the fast initial modes of spreading, but not with the slow propagating front. We point out that lateral spreading of activity is linked to pathological levels of persistent receptor activity as observed in cancer cells and exemplify uses of this link for the design and quick evaluation of molecular therapies targeting primary activation of EGFR.

## Introduction

The ErbB family of receptor tyrosine kinases, consisting of the epidermal growth factor receptor (EGFR, also called HER1 or ErbB 1) and three other members (referred to as Her 2--4 or ErbB 2--4), plays a prominent role in signaling pathways, which regulate important biological functions like proliferation, differentiation and cell survival. Dysfunctions and overexpression of these receptors are involved in carcogenesis. Since the seminal work of Kholodenko et al. [1] on the early stages of the EGF signaling cascade a number of increasingly refined and increasingly complex models of the ErbB signaling network have been published [2-4]. However, the very first stages of the cascade, especially the precise nature of transmembrane signaling after ligand binding remains poorly understood, despite the fact that it constitutes the most important target of molecular therapies in cancer cells overexpressing EGFR [5]. It is usually modeled as ligand-induced dimerization of receptors, followed by the activation of the tyrosine kinase and trans-phosphorylation of tyrosine residues on the C terminal. Unlike most other growth factor receptors, dimerization of EGFR is not achieved by crosslinking two receptor molecules with a bivalent ligand, and many experiments have revealed that primary EGFR activation is itself a highly complex and highly regulated process (see [6, 7]). The present body of experimental knowledge on structure and function of EGF receptors is not sufficient to offer a natural explanation for experiments, which demonstrate that EGFR phosphorylation can spread into EGF-free regions [8-10]. Verveer et al. [8] induced locally restricted activation of EGFR (fused with GFP) in the membrane of MCF7 cells by applying EGF covalently fixed to beads. Receptor phosphorylation was observed by measuring FRET between EGFR-GFP and a Cy3-labeled antibody to phosphotyrosine. They found lateral spreading of EGFR phosphorylation and they stressed the importance of secondary dimerization for this phenomenon. Sawano et al. [9] used a microfluidic device for local stimulation and report that the onset of lateral activity spreading depends on the concentration of EGF receptors in the plasma membrane. COS cells expressing EGF receptors at a normal density did not show spreading of activity. Reynolds et al. [10] studied a coupling of receptor kinase activity to an inhibition of protein tyrosine phosphatase via hydrogen peroxide (produced rapidly after EGFR stimulation) and already summarized their experimental findings in a schematic reaction scheme that implies bistable receptor activity.

All the experiments strongly suggest that there exists an autocatalytic property of EGF receptors (called "C" further on), which is switched on in a dimer consisting of active receptors and which turns these receptor molecules into activators of other ligand-free receptors. According to our present knowledge of the EGF receptor, this property is necessary for activation of receptors in EGF-free regions, although we should point out that it is lacking an explanation based upon structural features of the receptor molecule at present. We will show that property "C" can be implemented in a theoretical model without introducing any new, unknown parameters, if its kinetics is rigidly coupled to that of receptor phosphorylation. We will show that it provides quantitative explanations of key results of [8-10].

Activity spreading of EGFR emerges from a generic bistability of activity. This same bistability will lead to a pathologically high, permanent activity of genetically intact receptors, as it is observed in several types of cancer cells [11]. This link opens up new possibilities for the design and the evaluation of molecular therapies, which counteract the permanent receptor activity in cancer cells overexpressing EGFR.

## Results and Discussion

**Model building for primary receptor activation**

Usually, EGF receptors are activated by ligand binding, but here, we concentrate on activity spreading in EGF-free regions; therefore we start from a reaction-diffusion (RD) model, which contains receptor monomers and dimers, but does not include ligands and ligand binding. Here, we concentrate on a single type of ErbB receptors (EGFR) for simplicity, the neccessary refinements to include ErbB 2--4 are straightforward and will be presented elsewhere. We take into account dimerization/dissociation and phosphorylation/dephosphorylation of receptors, and we rigidly couple the property "C" to phosphorylation, i.e. every genetically intact receptor is assumed to have property "C" switched on, if its C-terminal sites are phosphorylated and switched off if they are not phosphorylated. In the absence of any information on the kinetics of property "C", this is the simplest hypothesis, which does not contradict the body of experimental observations. As we are not interested in downstream signaling, we do not discriminate between different phosphorylation sites. Specific rate constants are introduced for dimerization (D), for dissociation (-D) and for phosphorylation (P), with default values $k_D = 0.01\, nM^{-1} s^{-1}$, $k_{-D} = 0.01\, s^{-1}$, $k_P = 0.03\, s^{-1}$, respectively. Dephosphorylation (-P) is modeled by a Michaelis-Menten kinetics with values for $v_{max} = 0.5\, nMs^{-1}$ and for $K = 30\, nM$ taken from recent in vivo experiments by Yudushkin et al. [12]. The kinetics is assumed to be independent of dimerization. Therefore, the Michaelis-Menten rate law $k_{-P} = v_{max} / (K + n_a)$ depends upon the total density of active receptors, $n_a$. Additional phosphatase inhibition can be taken into account by smaller values of $K$ at fixed $v_{max}/K$. All the default values of the rate law parameters have been taken from independent standard models and experiments [1, 12-14], (see supporting text S1 for details). RD equations are set up for the densities $n_1$, $n_0$ of active and inactive monomers and $m_{00}$, $m_{10}$ and $m_{11}$ of dimers containing zero, one or two active receptors. These densities change by diffusion (diffusion constant $D_1 = 0.1\, \mu m^2 s^{-1}$ for monomers and $D_2 = D_1/2$ for dimers) and by reactions with velocities, which are collected in Tab.1. Note that we have added rates $b_{1P}$ and $b_{2P}$, which correspond to spontaneous phosphorylation of monomers or dimers, respectively, and which lead to a basal level of activity, which we adapted to specific cell types. For some numerical experiments, we also use spatiotemporal profiles of these rates as a substitute for ligands to drive receptor activation. The crucial autocatalytic reaction steps of property "C" are depicted in Fig. 1. Ultimately, EGFR activity is downregulated via receptor internalization, which we do not take into account here, as our main results will be of relevance for timescales of up to 5 minutes.

Using the operators $R_a = \partial_t - D_a \nabla^2$, with $a=1$ for monomers and $a=2$ for dimers, the complete set of RD equations of our model takes on the compact form $R_1 n_\mu = v_\mu^D + v_\mu^P$ and

$R_2 m_{\mu\lambda} = w^D_{\mu\lambda} + w^P_{\mu\lambda}$ with $\mu,\lambda \in \{0,1\}$. The system is discussed in detail in the supporting text S1.

In the absence of diffusion, the five rate equations of our model lead to bistable behaviour of the receptor activity within a wide range of parameters. This bistability constitutes the basis of spatial spreading of activity. We have numerically solved the model consisting of 5 RD equations using standard finite element methods, but to gain more insight into underlying mechanisms, additional analytical discussions are helpful. First, it is convenient to use a different set of variables. Instead of $n_0, n_1, m_{00}, m_{10}, m_{11}$, we introduce the density of monomers $n = n_0 + n_1$, the density of dimers $m = m_{11} + m_{10} + m_{11}$, and the total density of active receptors $n_a = n_1 + m_{10} + 2m_{11}$ together with $n_1$ and $m_{10}$ to describe the state of the system. The density $n_a$ is proportional to (ratiometrically calibrated) FRET signals of the experiments [8, 9, 10], which probe the phosphorylation of tyrosine residues of EGFR. Thus $n_a$ will be the quantity of central interest in the following. The spatial averages of densities $n_1$ and $n_a$ have to obey the inequalities $\overline{n_1} \leq \overline{n}$ and $\overline{n_a} \leq \overline{c} - \overline{n} + \overline{n_1} - \overline{m_{10}}$ to ensure positive $\overline{n_0}$ and $\overline{m_{00}}$. $\overline{c}$, the spatial average of receptor concentration is also used to define the dimensionless fraction of active receptors, $n_A = \overline{n_a}/\overline{c}$.

After expressing the RD equations in the new set of variables, two simplifications become obvious, which help to understand the dynamics of the system. First, the equations for $n$ and $m$ decouple from the rest of the system and take on the simple form (note that $n + 2m$ is the total receptor density)

$$(\partial_t - D_1 \nabla^2)n = -2(\partial_t - D_2 \nabla^2)m = k_{-D}m - k_D n^2.$$

The dimer/monomer equilibrium is not disturbed by changes in receptor activity, therefore we can use the stationary values $n^* = k_{-D}\left\{\sqrt{1 + 4k_D \overline{c}/k_{-D}} - 1\right\}/(2k_D)$ and $m^* = (\overline{c} - n^*)/2$. This simplification is due to our assumption of state-independent dimerization and dissociation rate constants. Second, the system can be approximated by a single equation for $n_a$, and this approximation, which is based on separation of timescales and lengthscales (relaxation of $n_a$ is much slower than that of the other densities), keeps the key aspects of the full dynamics. We introduce it here because we will make use of its explanatory power. Details of the model reduction can be found in the supporting text S1. The resulting RD equation for $n_a$ takes on the form

$$\partial_t n_a = f(n_a) + D_2 \nabla^2 (n_a - n_1^*(n_a)) + D_1 \nabla^2 n_1^*(n_a),$$

where $n_1^*$ is obtained by putting $(\partial_t - D_1 \nabla^2) n_1^* \approx 0$. The 2 diffusion terms correspond to the subpopulations of active dimers and active monomers contained in $n_a$.

**The dependence of activity spreading on receptor density is correctly predicted by the model**

Fig. 2 displays $f(n_a)$ for $\bar{c}=100\ nM$ and $\bar{c}=200\ nM$ and clearly exhibits a bistability of $n_a$ for an intermediate range of receptor densities and default values of kinetic parameters. In the bistable regime the system may be driven from a state of low, basal activity to a highly active state by the application of a stimulus, which can be localized in space and time as will be discussed in the next subsection. Indeed, a density dependence of spreading was observed in [9]. Therefore, as shown in Fig. 3, we have compared the density dependence of stationary and homogeneous activities resulting from our model to those of studied cell lines. According to our model, native COS cells ($\bar{c} \approx 20\ nM$) should not exhibit bistable behaviour and activity spreading, whereas COS cells overexpressing EGF receptors and MCF7 cells ($\bar{c} \approx 300\ nM$) should. This is in accordance with the results of [8-10]. A431 cells ($\bar{c} \approx 600\ nM$) should be constitutively active at pathologically high levels (even without permanently active mutants). Here, we have considered levels of basal activity of 3% - 5%. The lower the basal activity, the more the bistable regime extends towards higher densities.

**Activity spreading is based on a new generic bistable reaction scheme**

The reaction mechanism underlying this bistability is a robust and generic one, and is composed of an interplay of two steps, which are frequently met in biochemical reactions of the cell, namely, a deactivation via an enzymatic reaction, which is counteracted by an activation via dimerization. Fig. 4 shows a rate balance plot based upon a decomposition of the rate law of active receptors,

$$f(n_a) = r_+(n_a)n_a - k_{-P}(n_a)n_a$$

into an effective creation rate $r_+(n_a) = \{k_P m_{10}^*(n_a) + b_{1P}(n^* - n_1^*(n_a)) + b_{2P} m_{00}^*(n_a)\}/n_a$ and a destruction rate, which is just $k_{-P}(n_a)$, as expected. The structure of $r_+$ is also quite comprehensible: additional active receptors arise from auto-phosphorylation in dimers and from the spontaneous phosphorylation events in inactive monomers and inactive dimers. Note that for $b_{1P} = b_{2P} = 0$ there will be no receptor activation without dimerisation. The solid line in Fig. 4 has the well-known hyperbolic form of Michaelis-Menten kinetics. If it was combined with a creation rate, which increases with $n_a$, it would always lead to a single equilibrium point. The rate $r_+$, however, decreases with $n_a$ and gives rise to an extended region of bistability in parameter space. The steep decrease of $r_+$ for small $n_a$ is entirely due to spontaneous phosphorylation. The dash dotted curve corresponds to $r_+$ at $b_{1P} = b_{2P} = 0$ and is very close to a straight line, the hallmark of dimerization rates. Note that in this case $n_a = 0$ is always a fixpoint although it is not visible in the rate balance plot. We cannot extend the rates to $n_A = 1$, because values of $n_A$ beyond $0.93$ are not admissible in the reduced model as they violate the condition of positivity of $m_{00}$ discussed above.

**Experimental results on spreading of activity are explained quantitatively**

Let us now compare our model to the quantitative experimental results on spreading of activity in [8-10]. A suggestive first explanation of the observed spreading might identify it with a propagating front solution of the model. Standard analysis and numerical solution reveal that the full and the reduced model possess propagating plane front solutions of the form $n_a(x,y,t) = f(x-vt)$ on an extended rectangular membrane located in the $(x-y)$ plane [15]. The fronts propagate in positive x direction if receptors are kept active at the left boundary of the rectangle. The profile of the front is a sigmoidal function joining a high value $n_{a,l}$ at the left boundary to a low value $n_{a,r}$ at the right boundary and we may define a width of this profile by $w:=(n_{a,l}-n_{a,r})/|s|$, where $s$ is the slope of $f$ at its inflection point. We solved the full and the reduced model numerically on a long rectangular strip and studied the front connecting the two stable homogeneous fixpoints of our model for default kinetic parameters and $\bar{c}=200 nM$. We found a velocity of $v=0.014 \mu m s^{-1}$ and a width of $w=25.75 \mu m$. At first sight, these values seem to rule out that the observed spreading of activity is explainable by the asymptotic motion of the front solution, because the velocity is much slower than the observed spreading and the width of the front is comparable to the cell size.

Nevertheless, the model is in accordance with experimental findings, if we perform a more detailed comparison, including the initial, transient dynamics of a stimulated system, which can be much faster than the asymptotic propagating front. Let us first consider the experimental setups of [8] and [10], which use beads with covalently fixed EGF to stimulate the system. Interactions between ligand bound receptors and inactive receptors can only take place at the boundary of the stimulated regions. The beads are mapped to our model as circles (of $0.8 \mu m$ in diameter) with fixed boundary conditions of high receptor activity. In [8] cells on coverslips were drained and overlaid with a suspension containing $4 \times 10^5$ beads per microliter, which leads to a fluctuating number of beads fixed to the plasma membrane. From data of [8] and [10], we found $3\ldots20$ beads randomly distributed on the cell surface. Receptor activity is observed to spread over the surface on a timescale of at least $1\ldots2$ minutes. The average distance between beads depends upon the geometry of the cell attached to substrate. If we take a roughly cylindrical geometry of a cell with $1000\ \mu m^2$ surface, there is $500\ldots700\ \mu m^2$ of available membrane surface for stimulation, so that beads control areas of $2.8\ldots8.6\ \mu m$ radius. From the boundary of beads activity has to spread $2.5\ldots8\ \mu m$ before the activated regions start to overlap. Given the fluctuating number of beads we decided to study the activity spreading emerging from a single bead by following the temporal development of levels of fixed activity $n_a(r,t)/\bar{c} = const$ (activity isoclines) from a single bead. Fig. 5 reveals that initially levels of activity up to $0.4\ldots0.5$ spread several times faster than the asymptotic front and may overlap within 1--2 minutes. As soon as levels overlap, their development is accelerated further as shown in Fig. 6, which displays activity isoclines after 20, 40 and 60 seconds. On the other hand, high levels of activity spread significantly slower than the asymptotic front, so that receptor internalization sets in before these levels start to overlap.

Now we turn to [9], where ligand is applied in a microfluidic flow channel. This provides a ligand profile with sharp boundary, but in contrast to [8] and [10], ligand-bound EGFR molecules are free to diffuse. As the EGF unbinding rate is small, $0.004\ s^{-1}$, a receptor can diffuse for $250\ s$ before it looses its EGF ligand. It is important to include this smearing out effect of initial activity to achieve a satisfactory agreement between the model and experimental data. Fig. 7 compares numerical results from our model with Fig. 6 of [9], which shows the spreading of phosphorylation for transfected COS cells. We monitor the time course of activation $n_a$ at $x=0\ \mu m$, $x=10\ \mu m$, $x=20\ \mu m$, and $x=30\ \mu m$, which corresponds to the regions of interest marked in Fig 6. of [9].

The experimental setups may also be used to test an important prediction of the model, the appearance of critical pulses. For each spatial profile of activation the model predicts a critical minimal activation time necessary to induce activity spreading. For strip-like stimulation profiles as they are available in microfluidic devices, we have calculated that critical pulse duration is between $10\,s$ to $15\,s$ (see supporting text S2 for more details), if 10% of the receptors are stimulated.

**The model provides hints for targeted molecular therapies**

In cancer cells, which overexpress EGFR, tyrosine kinase inhibitors like gefitinib (Iressa$^{TM}$) and erlotinib (Tarceva$^{TM}$), and monoclonal antibodies against the EGFR ectodomain, like cetuximab (Erbitux$^{TM}$, for ErbB1), trastuzumab (Herceptin$^{TM}$, for ErbB2) have shown to suppress tumor growth with some success [16, 17]. Recently, the EGFR related protein (ERPP) has been shown to effectively inhibit several ErbB receptors including ErB1 and ErbB2, presumably by ligand binding and sequestration and by the formation of inactive heterodimers with ErbB receptors [18]. The inherent bistability of receptor activity discussed above provides a quantitative link between the phenomenon of activity spreading and the appearance of pathologically high levels of intact, persistently active receptors. The proposed model of primary activation helps in the design and quick evaluation of molecular therapies with targets, which directly influence parameters of the model.

To illustrate this point, we consider the control of phosphorylation rate $k_P$, a model parameter susceptible to kinase inhibitors, and of the total density of receptors $\bar{c}$, which in effect may be reduced by monoclonal antibodies, if they completely inactivate receptors. In Fig. 8, the stability regions in the $c_T - k_P$ plane are shown, other model parameters are kept to their default values. It illustrates the scenario of a combined therapy, which will lead back to normal basal activity level, although this cannot be achieved by separately applying a kinase inhibitor or an antibody, which inactivates receptors.

Fig. 9 displays the onset of pathological activity levels due to genetic defects, which increase $b_{1P}$ (as a first, oversimplified means to include effects of constitutively active $\Delta 2-7$ mutants) and overexpress EGFR. Note that attempts to reconstitute a stable low level of activity by reducing the spontaneous activation of monomers will only succeed in the small shaded region of Fig. 9, if it is not accompanied by a reduction of expression level. We have also studied other potential targets to inhibit EGFR activity and found that stabilizing dimers by reducing the dissociation rate $k_{-D}$ may also be a promising means to reestablish a normal activity level (see supporting figure S3 for more details). These simplified examples can easily be extended to more realistic settings, but they already show the potential for additional, quantitative methods of investigation in the field of EGFR targeted molecular therapies.

In summary, we have set up a new computational model of spatially resolved activation of EGFR, which explains experimental findings of lateral activity spreading. Our basic hypothesis, i.e. an autocatalytic activation kinetically linked to phosphorylation, proved to be in quantitative accordance with existing experiments so far, without introducing any new and unknown kinetic parameters. The activity spreading turned out to be based on a generic bistability, which will show up, whenever Michaelis Menten kinetics is counteracted by dimerization or other types of reactions possessing rates, which decrease with concentration of the product. Using kinetic constants from the established models we find that the bistability only occurs within an intermediate range of receptor densities. Our model correctly predicts, which of the experimentally used cell lines show activity spreading and which show homogeneous, low or high basal activity. The model allows to calculate the evolution of activity patterns following a stimulating pulse of any shape, strength and duration and thereby provides a wealth of experimentally testable predictions. We have shown in detail how to compare it to experimental findings.

It is an interesting feature of the model that it contains a quantitative link between the pathologically high activity levels in cancer cells overexpressing EGFR and the phenomenon of activity spreading. We have presented some oversimplified examples to demonstrate the usability of this link in the design of molecular therapies and to show that it can open up a new and promising field of quantitative research for molecular therapies.

Although our results indicate that bistability and activity spreading only occur in cell lines overexpressing EGFR, it may still be possible that the signal amplification due to activity spreading is a useful physiological mechanism. EGF receptors are known to cluster and can form patches of higher density, where activity spreading becomes possible. If spreading activity pulses are rare events and the cell can downregulate the receptor activity (via internalization of receptors), bistability constitutes a useful intrinsic discriminator between less important (sub-threshold) signals and important ones.

## Methods

The model of primary EGFR activation discussed above was formulated as 5 RD equations on bounded two-dimensional regions, either rectangluar or circular, with von Neumann boundary conditions. The model equations together with reformulations and reductions are discussed in detail in the supporting text S1. The system of parabolic partial differential equations was solved numerically by triangulating the reaction regions and using a standard, freely available finite element library [19] to solve the boundary value problems. The mesh size was chosen such that every cell constitutes a well-stirred reaction volume. The time evolution was constructed by discretising time and applying a simple Rosenbrock method [20]. Numerical errors were checked by changing discretisation parameters (time intervals and mesh sizes). The system was driven either by boundary conditions on additional interior boundaries representing localized ligand or by spatiotemporal profiles of spontaneous phosphorylation rates to model ligand pulses in microfluidic devices.

## Supporting material

Supporting material is available from
http://www.theorie.physik.uni-goettingen.de/~jentsch/supp.pdf

Thesis, Humboldt-Universität zu Berlin, Mathematisch-Naturwissenschaftliche Fakultät I).

15. Murray J.D. (1989) Mathematical Biology (Springer-Verlag Berlin, Heidelberg).

16. Paez J.G., Janne P.A., Lee J.C., Tracy S., Greulich H. et al. (2004) EGFR mutations in lung cancer: correlation with clinical response to gefitinib therapy. Science 304:1497-1500.

17. Yip Y.L., Ward R.L. (2002) Anti-ErbB2 monoclonal antibodies and ErbB-2-directed vaccines. Cancer Immunol Immunother 50: 569-587.

18. Xu H., Yu, Y., Marciniak, D, Rishi A.K., Sarkar, F.H. et. al. (2005) Epidermal growth factor receptor (EGFR)- related protein inhibits multiple members of the EGFR family in colon and breast cancer cells. Mol Cancer Ther 4: 435-442.

19. Hecht F., Pironneau O., Le Hyaric A., Ohtsuka K., (2005) FreeFem++, Available: http://www.Freefem.org.

20. Press W.H., Teukolsky S.A. Vetterling W.T. Flannery B.P. (2007) Numerical recipes 3rd edition: The art of scientific computing (Cambridge University Press)
**Figure 1**: **EGFR activation.** Left part: The standard ligand induced activation of EGFR. Ligand (black circle) binds to 2 receptor monomers with inactive kinase (white ellipse), which subsequently dimerize and thereby activate their kinases (black ellipse). All reaction steps

(undirected lines) are reversible. Right part: Activation of EGFR with property C. An active monomer may bind to an inactive one. The resulting dimer can either switch to an inactive state (irreversibly, as indicated by the directed line), thereby enhancing the population of inactive receptors or the active receptor can (via property C) activate the other receptor. After dissociation, this enhances the population of active monomers with property C

**Figure 2**: **Bistability of EGFR activity.** Reaction velocity of active receptors $f(n_a)$ vs. the fraction of active receptors $n_A = n_a / \bar{c}$ for average receptor densities of $\bar{c}=200 nM$ (upper curve) and $100 nM$ (lower curve) and for default kinetic parameters. Zeroes of $f$ correspond to homogeneous, stationary states. Spatially homogeneous initial conditions with $n_A$ beyond the threshold (2) are driven to the fixpoint with high activity.

**Figure 3: Dependence of fixpoints on total receptor concentration.** Fixpoints of $n_A$ vs. average receptor concentration $c_T = \bar{c}$. The vertical lines a), b) and c) correspond to concentrations of normals COS cells (a), transfected COS cells overexpressing EGFR as used in Ref. [9] (MCF7 cells as used in Ref. [8] have comparable densities) and (c) corresponds to A431 cells. We estimate the densities for a cell with a radius of $10 \mu m$, where $c_T = 100 nM$ corresponds to $2.4 \times 10^5$ receptors.

**Figure 4: Rate balance plot.** Solid line: Michaelis-Menten rate law of dephosphorylation. Dotted line: effective activation rate $r_+$, which contains spontaneous phosphorylation (dominating for $n_a \to 0$) and a rate limiting dimerization step. Dash-dotted line: $r_+$ without spontaneous phosphorylation. The reduced model correctly reproduces the equilibrium states per definition, but looses its applicability in the gray shaded area, where it violates the positivity of $m_{10}$.

**Figure 5: Activity spreading from focal stimulation.** Activity isoclines of a single bead are circles of radius $r(t)$ increasing with time. An average (radial) velocity up to time $T$ is defined as $v(T) = (r(T) - r(0))/T$. The figure shows $v(60sec)$ and $v(300sec)$ in units of the front velocity $v_{front} = 0.014 \mu m s^{-1}$ vs. the activity level. Lower activity levels start out with a considerably higher velocity and slow down, but the velocities of higher activity levels may fall well below $v_{front}$.

**Figure 6: Activity spreading from focal stimulation with two beads.** The $n_a = 0.1$ isoclines of two beads (shaded circles) are shown for $T=20, 40$ and $60$ seconds. Note that the separation between the isoclines first decreases, but increases again when the activated regions from both beads start to overlap.

**Figure 7: Modelling microfluidic stimulation.** To compare with experimental results from Ref. [9] we solve the RD equations on a $30 \times 20 \mu m$ membrane with a strip-like, sigmoidal activation $b_{1P}(x) = b_{1P} \tanh(x/\xi)$ with $\xi = 13 \mu m$. In a homogeneous stimulation, $b_{1P}$ will lead to a stationary activation of $n_a \approx 0.93$. The curves correspond to different distances from the left boundary ($x = 0, 10, 20, 30 \mu m$), chosen in accordance with the regions of interest monitored in [9]. The experimental results consist of normalized FRET signals and are shown in the inset. See [9] for further details.

**Figure 8: Phase diagram in the $k_P - c_T$ plane.** The solid lines divide the $k_P - c_T$ plane into states of pathologically high basal activity (upper part), bistable activity (as marked, considered still as pathological activity here) and states of normal basal activity (lower part) (phase diagram). Therapies, which reduce the expression level ($c_T$) of receptors or inhibit kinase activity by certain values, may not be completely successful (as indicated by arrows). When both therapies have failed, a combination of the two therapies may lead back to normal basal activity.

**Figure 9: Phase diagram in the $b_{1P} - c_T$ plane.** An increase of spontaneously active receptors may be due to a genetic defect. Note that there is only a very small regime of $c_T$ expression levels (gray shaded), where reducing the level of genetically defect receptors is an efficient therapy.

**Table 1:** Reaction velocities

| velocity | decrease | increase |
|---|---|---|
| $v_0^D$ | $-k_D n_0^2 - k_D n_0 n_1$ | $+k_{-D}(2m_{00} + m_{10})$ |
| $v_0^P$ | $-b_{1P} n_0$ | $+k_{-P}(n_a) n_1$ |
| $v_1^D$ | $-k_D n_1^2 - k_D n_0 n_1$ | $-k_D(2m_{11} + m_{10})$ |
| $v_1^P$ | $-k_{-P}(n_a) n_1$ | $+b_{1P} n_0$ |
| $w_{00}^D$ | $-k_{-D} m_{00}$ | $+k_D n_0^2 / 2$ |
| $w_{00}^P$ | $-2b_{2P} m_{00}$ | $+k_{-P}(n_a) m_{10}$ |
| $w_{10}^D$ | $-k_{-D} m_{10}$ | $+k_D n_0 n_1$ |
| $w_{10}^P$ | $-k_P m_{10} - k_{-P}(n_a) m_{10}$ | $+2k_{-P}(n_a) m_{11} + 2b_{2P} m_{00}$ |
| $w_{11}^D$ | $-k_{-D} m_{11}$ | $+k_D n_1^2 / 2$ |
| $w_{11}^P$ | $-2k_{-P}(n_a) m_{11}$ | $k_P m_{10}$ |

Reaction velocities of monomers are denoted by v (those of dimers by w). The lower index refers to the subpopulations (0=inactive, 1=active) and the upper index characterizes the reaction (P=(de)-phosphorylation, D=dimerisation/dissociation). The velocities are decomposed into parts, which increase or decrease the corresponding subpopulation.

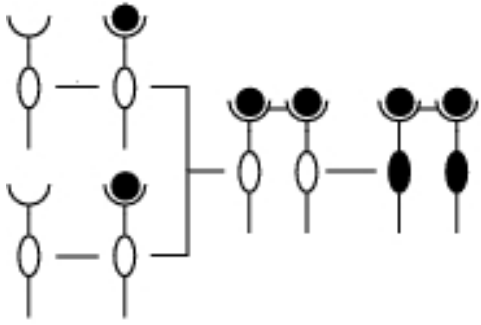 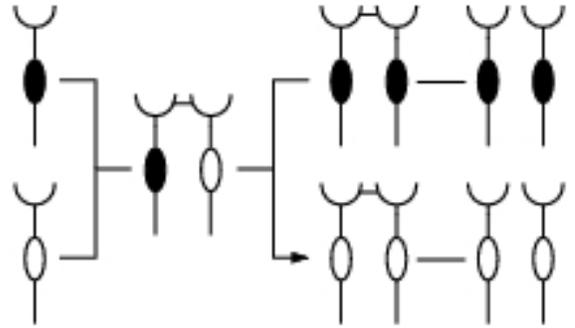

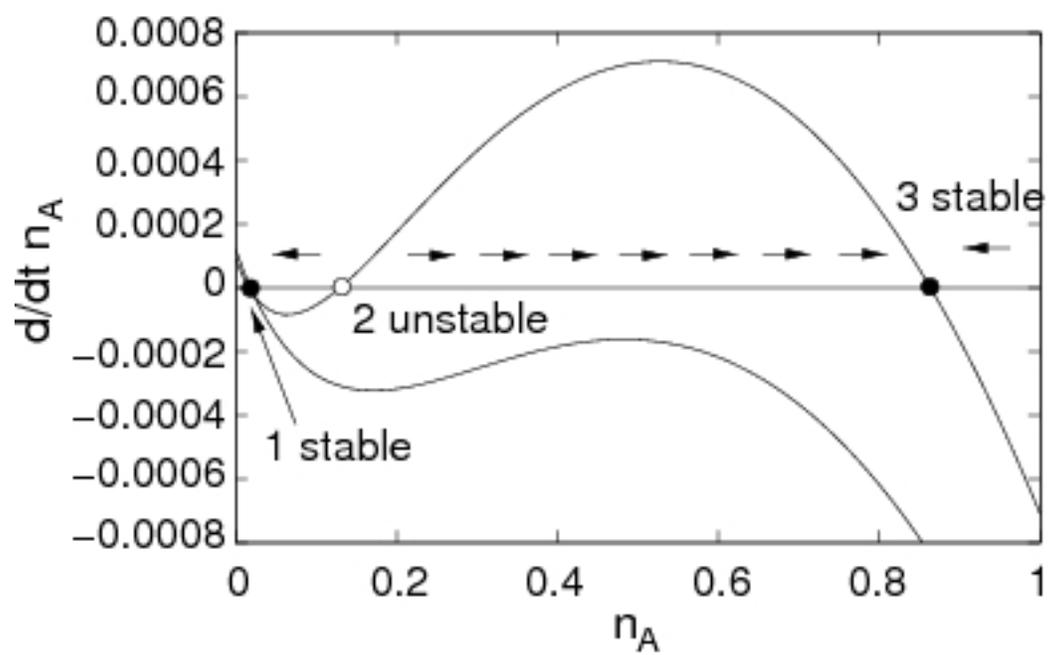

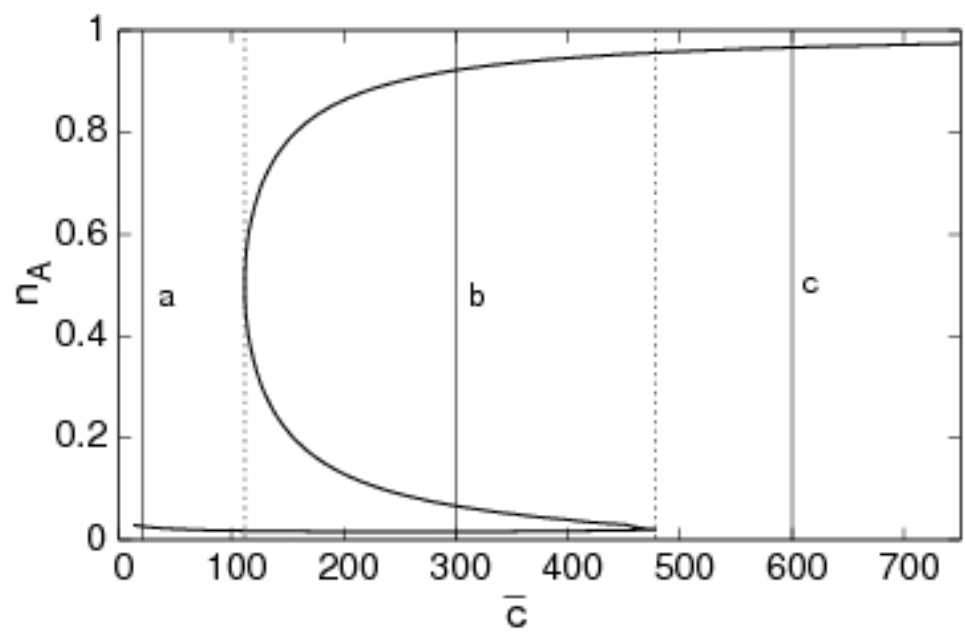

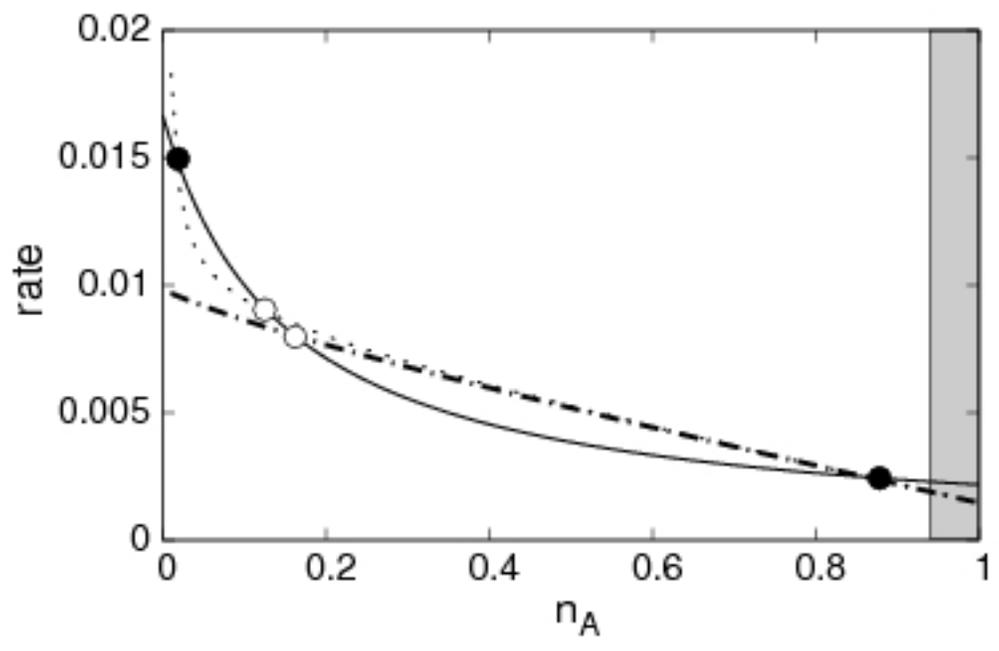

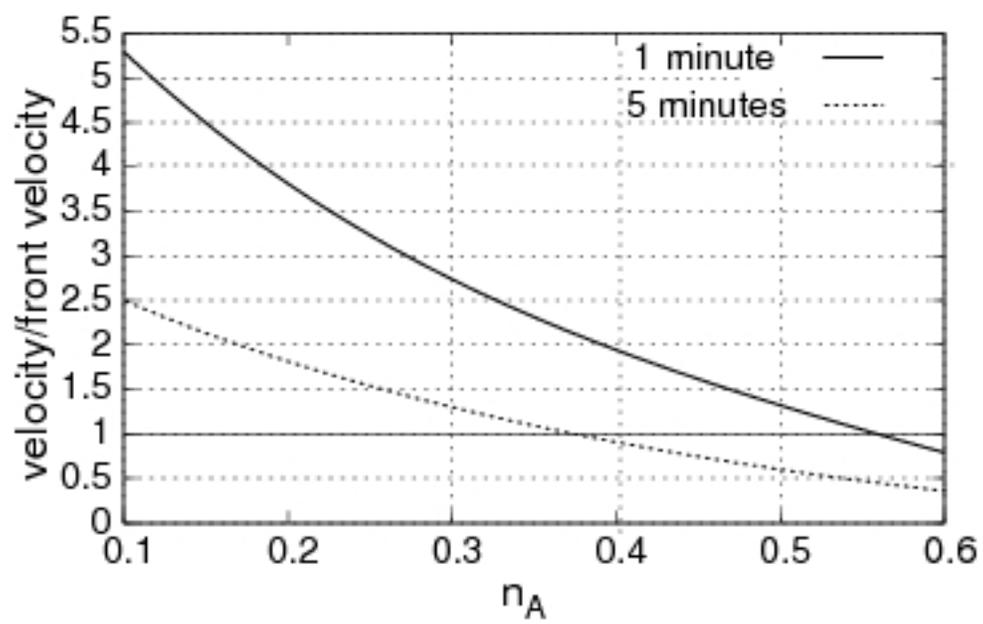

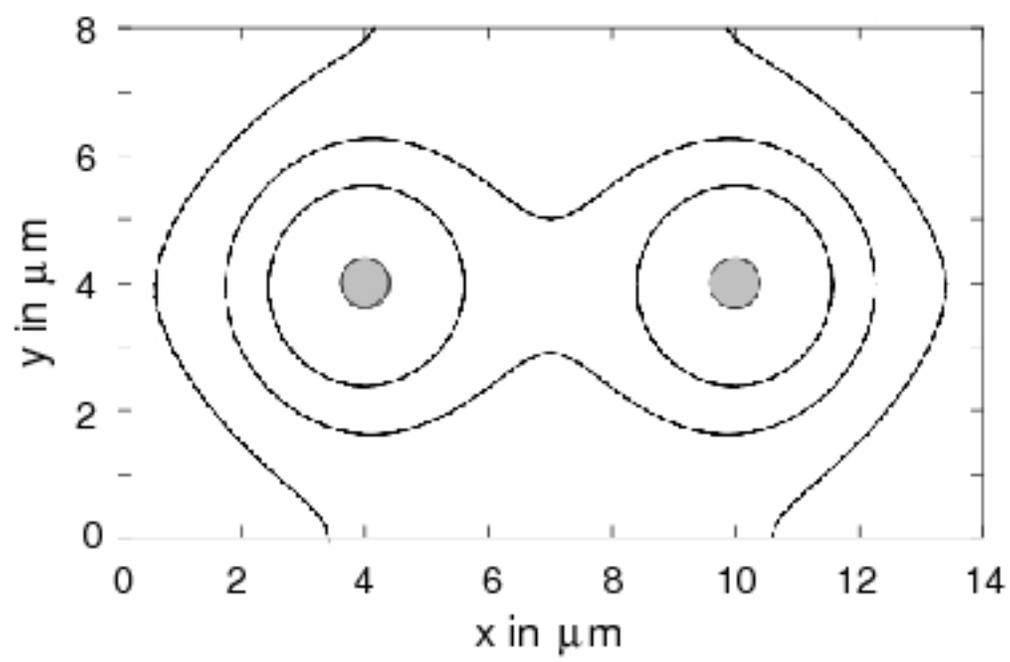

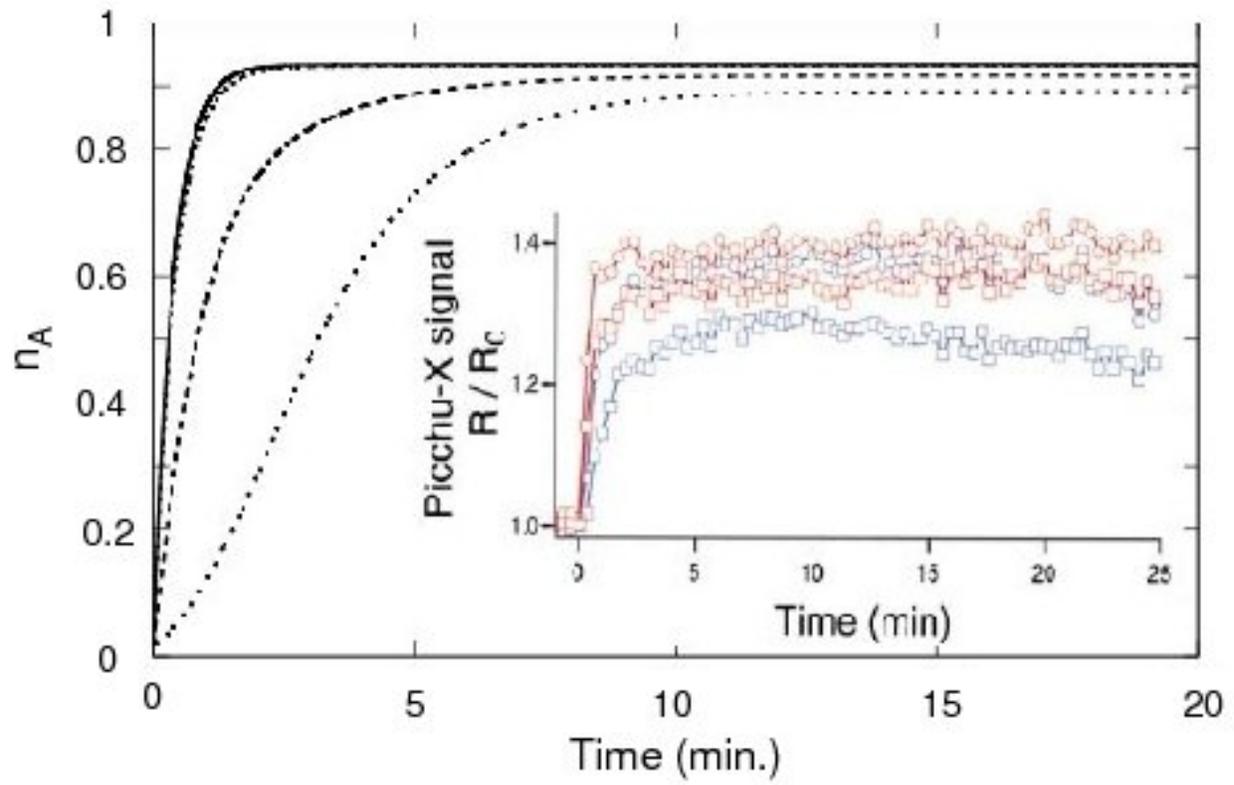

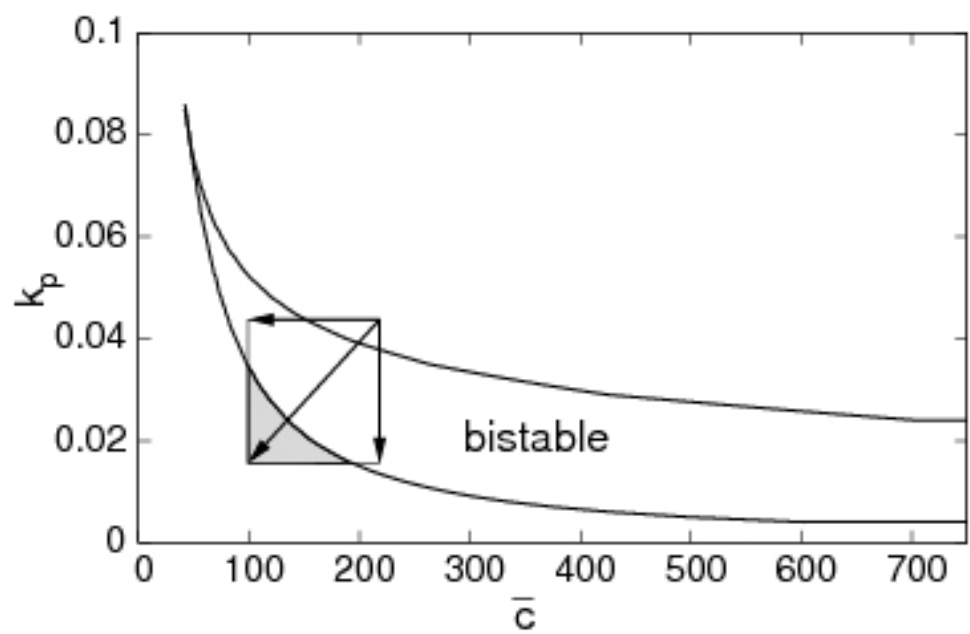

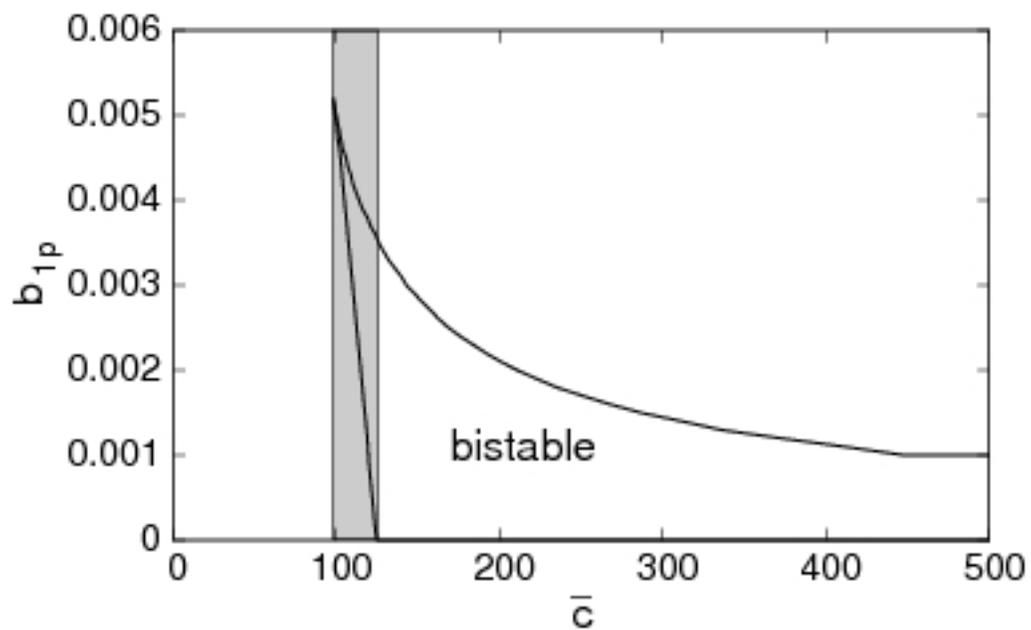